\PassOptionsToPackage{table,usenames,dvipsnames}{xcolor}
\documentclass[10pt,a4paper,hidelinks]{article}
\newcommand{\mytitle}{Inferring Input Grammars from~Dynamic~Control~Flow}

\usepackage{fancyhdr}
\usepackage{pdfpages}
\usepackage[procnames]{listings}
\usepackage{graphicx}

\fancypagestyle{CISPA}{

\fancyhead[C]{\includegraphics[width=4cm]{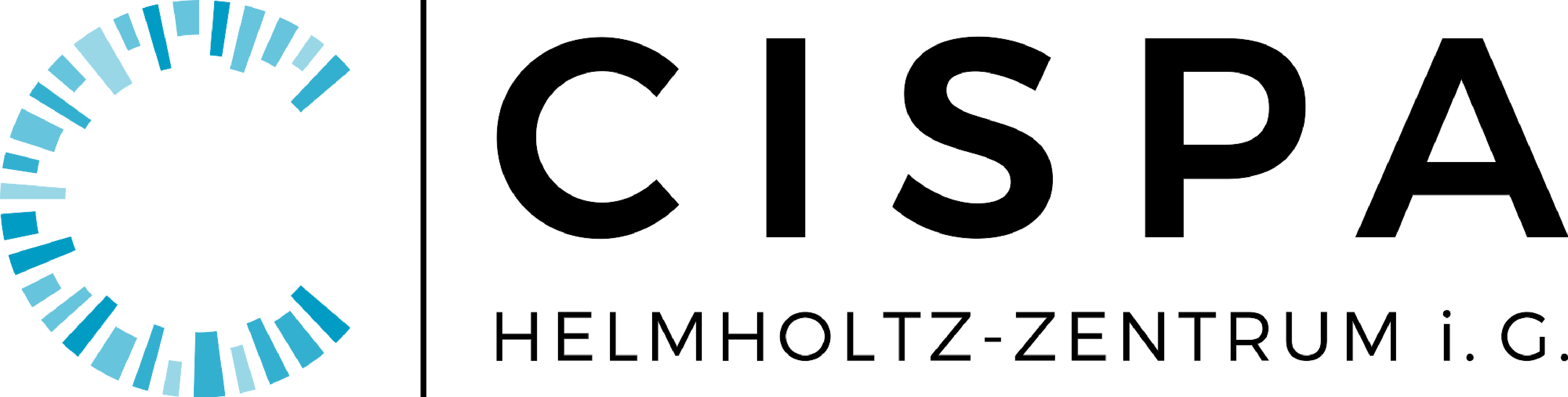}}
}
\usepackage{etoolbox}
\usepackage[backend=biber,bibencoding=utf8]{biblatex}
\usepackage{etoolbox}
\newif\ifdraft\draftfalse
\newif\iflong\longtrue

\usepackage[T1]{fontenc}
\usepackage[utf8]{inputenc}
\usepackage{csquotes}
\usepackage[english]{babel}
\usepackage[nohead,includefoot,margin=2cm]{geometry}
\usepackage[compact,raggedright,small]{titlesec}
\usepackage[affil-it]{authblk}
\usepackage[runin]{abstract}
\usepackage{multicol}
\usepackage{titling}
\usepackage{montserrat}
\usepackage{setspace}
\usepackage{booktabs}   %
\usepackage{multirow}
\usepackage{subcaption} %
\usepackage{caption}

\usepackage{hyperref}
\hypersetup{colorlinks=true,citecolor=blue,linkcolor=blue,filecolor=blue,urlcolor=blue,pdftitle={\mytitle}}
\usepackage[inline]{enumitem}
\usepackage{microtype} %
\usepackage{alltt}
\usepackage{float}
\usepackage{stfloats}

\usepackage[noend]{algpseudocode}
\usepackage{color}
\usepackage{relsize} %

\usepackage{comment}
\usepackage{amsmath,amssymb,amsfonts}
\usepackage{algorithm}
\usepackage{textcomp}
\usepackage{xspace}
\usepackage{syntax}
\usepackage{qtree}
\ifdraft
\usepackage{todonotes}
\else
\usepackage[disable]{todonotes}
\fi
\usepackage{multirow}
\usepackage{tabularx}

\usepackage[nameinlink,noabbrev]{cleveref}  %

\definecolor{codegreen}{rgb}{0,0.6,0}
\definecolor{codegray}{rgb}{0.5,0.5,0.5}
\definecolor{codepurple}{rgb}{0.58,0,0.82}
\definecolor{backcolour}{rgb}{0.95,0.95,0.92}

\lstdefinestyle{mystyle}{
    commentstyle=\color{codegreen},
    keywordstyle=\color{magenta},
    numberstyle=\tiny\color{codegray},
    stringstyle=\color{codepurple},
    basicstyle=\linespread{0.8}\footnotesize\ttfamily,
    breakatwhitespace=false,
    breaklines=true,
    captionpos=b,
    keepspaces=true,
    numbers=left,
    numbersep=5pt,
    showspaces=false,
    showstringspaces=false,
    showtabs=false,
    tabsize=2
}

\algblockdefx[switch]{Switch}{EndSwitch}%
[1]{\textbf{switch} #1}%
{}%
\algloopdefx{Case}[1]{\textbf{case} #1:}%
\algdef{SE}[DOWHILE]{Do}{doWhile}{\algorithmicdo}[1]{\algorithmicwhile\ #1}%

\def\|#1|{\textit{#1}}
\def\<#1>{\texttt{#1}}

\lstdefinestyle{grammar}
{
    basicstyle=\linespread{0.8}\footnotesize\ttfamily,
    keywordstyle=\bfseries,
    numberblanklines=false,
    language=ebnf,
    tabsize=2,
    keywordstyle=\color{blue},
    identifierstyle=\color{red},
}

\definecolor{eclipseBlue}{RGB}{42,0.0,255}
\definecolor{eclipseGreen}{RGB}{63,127,95}
\definecolor{eclipsePurple}{RGB}{127,0,85}

\lstdefinestyle{python}
{
    basicstyle=\footnotesize\ttfamily,
    numberblanklines=false,
    language=python,
    tabsize=2,
    commentstyle=\color{eclipseGreen},
    keywordstyle=\bfseries\color{eclipsePurple},
    stringstyle=\color{eclipseBlue},
    procnamestyle=\bfseries\color{black},
    procnamekeys={def},
    columns=flexible,
    identifierstyle=
}

\def\BibTeX{{\rm B\kern-.05em{\sc i\kern-.025em b}\kern-.08em
    T\kern-.1667em\lower.7ex\hbox{E}\kern-.125emX}}

\newcommand{\AUTOGRAM}{\textit{Autogram}\xspace}
\newcommand{\Mimid}{\textit{Mimid}\xspace}
\newcommand{\GLADE}{\textit{GLADE}\xspace}
\newcommand{\REINAM}{\textit{REINAM}\xspace}
\newcommand{\GRIMOIRE}{\textit{GRIMOIRE}\xspace}
\newcommand{\mimid}{\textit{mimid}\xspace}
\newcommand{\nonterminal}{\textit{nonterminal}\xspace}
\newcommand{\nonterminals}{\textit{nonterminals}\xspace}

\newcommand{\sif}{\texttt{if}\xspace}
\newcommand{\selse}{\texttt{else}\xspace}

\newcommand{\sloop}{\texttt{loop}\xspace}

\newcommand{\pta}{\emph{PTA}\xspace}
\newcommand{\calc}{\emph{calc.py}\xspace}
\newcommand{\mathexpr}{\emph{mathexpr.py}\xspace}
\newcommand{\urlparse}{\emph{urlparse.py}\xspace}
\newcommand{\netrc}{\emph{netrc.py}\xspace}
\newcommand{\microjson}{\emph{microjson.py}\xspace}
\newcommand{\cgidecode}{\emph{cgidecode.py}\xspace}

\usepackage{framed}
\newenvironment{result}{\begin{framed}\centering\it}{\end{framed}}
\title{Inferring Input Grammars from~Dynamic~Control~Flow}    %
\date{\small (Dated \today)}
\author{Rahul Gopinath}
\author{Bj\"orn Mathis}
\author{Andreas Zeller}
\affil{\{rahul.gopinath, bjoern.mathis, zeller\}@cispa.saarland \\
CISPA - Helmholtz Center for Information Security, Saarbr\"ucken, Germany}

\newcommand\BackgroundPic{
    \put(0,0){
    \parbox[b][\paperheight]{\paperwidth}{%
    \vfill
    \centering
    \includegraphics[width=\paperwidth,height=\paperheight]{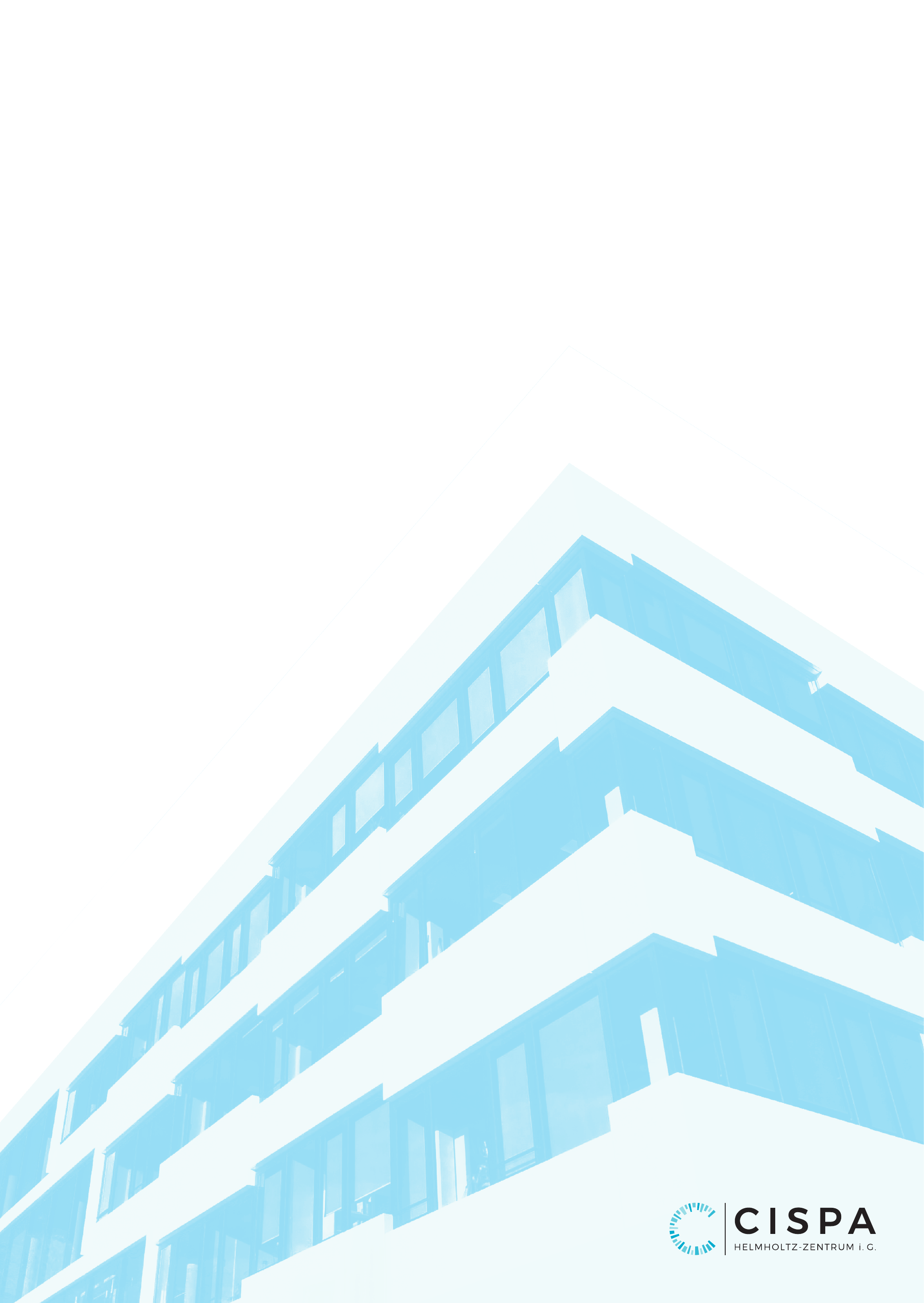}
    \vfill
}}}

\usepackage{utopia}
\usepackage{times}

\begin{document}
\AddToShipoutPicture*{\BackgroundPic}

\makeatletter
\renewcommand{\Authfont}{\normalsize\sffamily\bfseries}
\renewcommand{\Affilfont}{\normalsize\sffamily\mdseries}
\begin{titlepage}
\newcommand{\HRule}{\rule{\linewidth}{0.1mm}}
\centering
  \textsc{\LARGE {\fontfamily{Montserrat-TOsF}\selectfont CISPA Helmholtz-Zentrum i.G.}}\\[1.5cm]

  \vspace{2.4 cm}
  \HRule \\[0.2cm]
  {\huge\sffamily\bfseries \@title\par}
  \vspace{0.2cm}
  \HRule \\[1.5cm]

  {\sffamily \@author\par}
\vfill

\end{titlepage}

\makeatother
\setlength{\affilsep}{0.1em}
\addto{\Affilfont}{\small}
\renewcommand{\Authfont}{\normalsize}
\renewcommand{\Affilfont}{\normalsize}

\pretitle{\begin{center}\large\bfseries}
\posttitle{\end{center}}
\maketitle
\thispagestyle{CISPA}

\begin{abstract}
A program is characterized by its input model, and a formal input
model can be of use in diverse areas including vulnerability
analysis, reverse engineering, fuzzing and software testing, clone detection
and refactoring.
Unfortunately, input models for typical programs are often unavailable
or out of date. While there exist algorithms that can mine the syntactical
structure of program inputs, they either produce unwieldy and
incomprehensible grammars, or require heuristics that target specific parsing
patterns.

In this paper, we present a general algorithm that takes a program and a small
set of sample inputs and automatically infers a readable context-free grammar
capturing the input language of the program.  We infer the
syntactic input structure only by observing access of input characters at
different locations of the input parser.  This works on all program stack based
recursive descent input parsers, including PEG and parser combinators, and can
do entirely without program specific heuristics.  Our \Mimid prototype produced
accurate and readable grammars for a variety of evaluation subjects, including
expr, URLparse, and microJSON.
\end{abstract}
\begin{multicols}{2}

\section{Introduction}
\label{sec:introduction}
One of the key properties of a program is its input model, and the
availability of a formal input model can be critical in diverse fields.
\iflong
For example, an \emph{accurate} input model of a suspect program can indicate
the presence of hidden features.
When programs with similar input models are found, it is often a hint that
they may be clones and could be a target for refactoring (or may share similar
origin in the case of plagiarism detection).
The ability to generate valid
or near valid inputs for a program is also much sought after in software
testing, and especially fuzzing and vulnerability analysis~\cite{li2018fuzzing}.
Indeed, for
fuzzing to reach
deeper program states, one needs to be able to generate valid or near valid
inputs~\cite{valentin2018fuzzing}.\fi Unfortunately, generating valid inputs
is a non-trivial problem even when the source code is
available~\cite{godefroid2008grammar}.
In the vast majority of the cases, a formal input model may be
unavailable, and in the cases where an input model is available, it may not be
complete~\cite{bastani2017synthesizing}, or it may be obsolete, or inaccurate
with respect to the program~\cite{walkinshaw2007reverse}.
However, for testing complex inputs, a model for the input language is
practically mandatory~\cite{holler2012langfuzz,aschermann2019nautilus,godefroid2008grammar}.
\emph{Obtaining input models automatically} therefore bears great promise for
test generation~\cite{godefroid2008grammar,majumdar2007directed}, but also for
reverse engineering protocols~\cite{caballero2007polyglot}, program                                      refactoring~\cite{hauptmann2015generating},
and program comprehension~\cite{rajlich2002the,deprez2000a}.

A small number of tools exist that learn aspects of the input structure
in order to generate further inputs. \emph{Learn\&Fuzz} by
Godefroid et al.~\cite{godefroid2017learn} uses statistical models
as the generative representation. \GLADE and
\REINAM~\cite{bastani2017synthesizing,wu2019reinam} use internal structures that
take the shape of a grammar. \GRIMOIRE~\cite{blazytko2019usenix} learns partial
grammars, generalizing over input fragments that cover the same
code.  All these tools are focused on fuzzing, and can infer some aspects of the
input language, which is then used to generate inputs. But none of them claims
that the inferred \emph{intermediate} models would be readable, editable or
otherwise useful for humans in any way.

\begin{figure}[H]
\begin{grammar}
<START> ::= <json\_raw>

<json\_raw>  ::= `"' <json\_string\('\)> | `[' <json\_list\('\)> | `{' <json\_dict\('\)>
\alt <json\_number\('\)> | `true' | `false' | `null'

<json\_number\('\)> ::= <json\_number>+
\alt <json\_number>+ `e' <json\_number>+

<json\_number> ::= `+' | `-' | `.' | [0-9] | `E' | `e'

<json\_string\('\)> ::= <json\_string>* `"'

<json\_list\('\)> ::= `]'
  \alt <json_raw> (`,' <json_raw> )* `]'
  \alt ( `,' <json_raw> )+ (`,' <json_raw> )* `]'

<json\_dict\('\)> ::= `\}'
  \alt ( `"' <json\_string\('\)> `:' <json\_raw> `,' )* \\
  `"' <json\_string\('\)> `:' <json\_raw> `\}'

<json\_string>   ::=  ` ' | `!' | `#' | `\$' | `\%' | `&' | `''
\alt `*' | `+' | `-' | `,' | `.' | `/' | `:' | `;'
\alt `<' | `=' | `>' | `?' | `@' | `[' | `]' | `^' | '_', '`',
\alt `\{' | `|' | `\}' | `~'
\alt `[A-Za-z0-9]'
\alt `\textbackslash' <decode\_escape>

<decode\_escape> ::= `"' | `/' | `b' | `f' | `n' | `r' | `t'
\end{grammar}
\caption{JSON grammar extracted from \microjson.}
\label{fig:mimidjson}
\end{figure}

\newlength{\widest}%
\settowidth{\widest}{aaaaaaaaaaaaaaaaaaaaaaaaaaa  a}
\newcommand{\rhs}[1]{\makebox[\widest][l]{#1}}

The one approach so far that aims to produce human-readable and maintainable
descriptions of input structure
is \AUTOGRAM~\cite{hoschele2016mining,hoschele2017active} by H{\"o}schele et al.
Given a program and a set of inputs, \AUTOGRAM extracts a context-free grammar
that approximates the program's input language.  It does so by tracking the
\emph{dynamic data flow} between \emph{variables} at different locations in
the parser: If a substring of the input flows into a variable called \texttt{protocol},
this substring forms a \textit{protocol} \nonterminal in the grammar.

While \AUTOGRAM produces well-structured and readable grammars on a number of
subjects, it also depends on a number of assumptions, the most important being
that some data flow has to be there in the first place.  If a program accepts
a structured input where only part of the input is saved and used,
\AUTOGRAM has no data flow to learn from in the parts that were not saved.
Second, one of the traditional parsing techniques is to attempt to parse a
given string using a particular rule, and if the parse fail, attempt the next
rule in the sequence. For example, a parser may attempt to parse a rule
\texttt{if <body> then <body> else <body> end}, failing which, it may
try and succeed in parsing \texttt{if <body> then <body> end}.
This parsing strategy is also used by the PEG parsing
technique. In such a case, \AUTOGRAM~\cite{hoschele2017active} has multiple
conflicting data flows in that it now has to deal with the data flow in the
failing parse, but has no strategy to resolve the conflict.

Furthermore, \AUTOGRAM requires special heuristics to work
around common parsing patterns identified; the data flow induced by parser
lookahead, for instance, has to be ignored as it would otherwise break the
model~\cite{hoschele2016mining}.  Finally, common patterns such as passing
the complete input as an array with an index indicating current parse status
can break the subsumption model of \AUTOGRAM.
These limitations make the \AUTOGRAM approach hard to generalize for a wide
class of input processors.

In this paper, we describe a \emph{general algorithm} to recover the \emph{input
grammar} from a program without any of these limitations.  Rather than being
based on \emph{data flow,} it recovers the input grammar from
\emph{dynamic control flow} and how input characters are \emph{accessed} from
different locations in the parser.
Our algorithm works regardless of whether and how the parsed data is stored,
and requires no heuristics to identify parsing patterns.
It works on all program stack based recursive descent parsers, including PEG and
parser combinators; this parser class makes up 80\% of the top programming
language parsers on GitHub~\cite{mathis2019parser}.

The resulting grammars are well-structured and very readable.  As an example,
consider the JSON grammar shown in \Cref{fig:mimidjson}, which our \mimid prototype
extracted from \microjson{}.\footnote{We removed rules pertaining to whitespace
processing for clarity.}  Each JSON element has its own production rule;
$\langle$\emph{json\_number}$\rangle$, for instance, lists a number as a
string of digits.  Rules capture the recursive nature of the input: A
$\langle$\emph{json\_list}$'\rangle$ contains $\langle$\emph{json\_raw}$\rangle$
elements, which in turn are other JSON values.  All identifiers of nonterminals
are derived from the names of the input functions that consume them.  All this
makes for very readable grammars that can be easily understood, adapted,
and extended.

\begin{figure}[H]
\begin{subfigure}{\linewidth}
\begin{lstlisting}[style=mystyle, language=Python]
    def digit(i):
        return i in "0123456789"

    def parse_num(s,i):
        n = ''
        while i != len(s) and digit(s[i]):
            n += s[i]
            i = i +1
        return i,n

    def parse_paren(s, i):
        assert s[i] == '('
        i, v = parse_expr(s, i+1)
        if i == len(s): raise Ex(s, i)
        assert s[i] == ')'
        return i+1, v

    def parse_expr(s, i = 0):
        expr, is_op = [], True
        while i < len(s):
            c = s[i]
            if digit(c):
                if not is_op: raise Ex(s,i)
                i,num = parse_num(s,i)
                expr.append(num)
                is_op = False
            elif c in ['+', '-', '*', '/']:
                if is_op: raise Ex(s,i)
                expr.append(c)
                is_op, i = True, i + 1
            elif c == '(':
                if not is_op: raise Ex(s,i)
                i, cexpr = parse_paren(s, i)
                expr.append(cexpr)
                is_op = False
            elif c == ')': break
            else: raise Ex(s,i)
        if is_op: raise Ex(s,i)
        return i, expr

    def main(arg):
        return parse_expr(arg)
\end{lstlisting}
\caption{A Python parser for math expressions}
\label{fig:calcprogram}
\end{subfigure}
\vspace{\baselineskip}
\begin{subfigure}{\linewidth}
   \resizebox{.99\columnwidth}{.050\totalheight}{
\Tree[.parse\_expr
  [.while(1)
                      [.if(1)
                        [.parse\_num
                          [.digit
                        \textit{9}
                          ] ] ] ]
  [.while(1) \textit{+} ]
  [.while(1)
                      [.if(1)
                        [.parse\_num
                          [.digit
                        \textit{3}
                          ] ] ] ]
  [.while(1) \textit{/} ]
  [.while(1)
                      [.if(1)
                        [.parse\_num
                          [.digit
                        \textit{4}
                          ] ] ] ]
                          ]}
\caption{Derivation tree for \texttt{9+3/4}.}%
  \label{fig:derivationtree}
\end{subfigure}
\vspace{\baselineskip}
\begin{subfigure}{\linewidth}
\begin{grammar}
<START> ::= (<parse\_expr>"[*+-/]")*<parse\_expr>

<parse\_expr>  ::= <parse\_num> | <parse\_paren>

<parse\_paren> ::= `(' <parse\_expr> `)'

<parse\_num>   ::= <parse\_digit>+

<digit> ::= `0' | `1' | `2' | `3' | `4' | `5' | `6' | `7' | `8' | `9'
\end{grammar}
\caption{A grammar derived from the parser in \Cref{fig:calcprogram}}
\label{fig:calcgrammar}
\end{subfigure}

\caption{An expression parser, its parse tree, and the extracted grammar}
\label{fig:calc}
\end{figure}
\subsection{Why should one care about readable grammars?}

Grammars have a number of uses beyond simply fuzzing. These can be
used for 1) reverse engineering, 2) clone detection, 3) program comprehension,
4) documentation, 4) refactoring, 5) parsing, 6) runtime verification,
7) data transformation, 8) reducing inputs, 9) feature location,
10) automatic repair, and a number of other uses in software engineering.
Even if one focuses on
fuzzing, readable grammars allow practitioners to edit and augment them
to control what should be produced, for instance by providing probabilities or
adding constraints such as allowing only previously defined variables.

Readable grammars allow one to refine the grammar with specific
inputs such as logins, passwords, or exploits.
Given such a grammar, one can contract the grammar such that only specific
subparts be generated if one is first able to understand what parts of the
grammar correspond to the part that one is interested in.
If even a partial human readable grammar is available, it can be expanded
with human knowledge on features where the miner may not have sufficient inputs,
or identify vulnerabilities through human inspection of the grammar (e.g.
allowing special characters in usernames).
Fuzzers can only allow for such control if the model is human-readable.
Finally, we note that \AUTOGRAM and \mimid are the only tools available
right now that can actually recover a context-free grammar from a given
program. Other tools such as \GLADE, \REINAM, \GRIMOIRE and others only
recover a ``grammar like structure'' which are not translatable to a
structure consumable by standard off the shelf tools such as fuzzers and
parsers.

How does this work?  We use lightweight instrumentation to track
\emph{dynamic control flow} and use \emph{lightweight string wrappers} to
identify in which control flow nodes, specific input characters are accessed.
The character accesses as well as the corresponding control flow nodes are then
logged.
A \emph{parse tree} of the input string is extracted from that trace
using the following rules:

\begin{enumerate}
\item Names of methods that \emph{process}\footnote{
  In contrast to \AUTOGRAM which assumes that the name of the method that \emph{receives}
  a part of the input is the \nonterminal symbol for that part.
}
some part of the input is used as the \nonterminal symbol for the input
grammar for that part. As an example, consider \Cref{fig:calcprogram} showing a complete Python program to accept mathematical expressions.  The method \texttt{parse\_num()}, which parses numeric elements, becomes
the \nonterminal \texttt{parse\_num} in the parse tree (\Cref{fig:derivationtree}), representing numeric elements in
the input.

\item Parsers typically do not reparse their input if the
last parse was successful. That is, the method that accesses a particular input
character last, \emph{consumes} that character. If \texttt{digit()} is
the last to access the digit \texttt{3}, then \texttt{3} is consumed
by \texttt{digit()}.

\item Characters consumed in a method during each method call become alternative
expansions for the corresponding \nonterminal. \texttt{parse\_num()} will see
several different sequences of digits (\Cref{fig:derivationtree}), all forming
alternatives.

\item If some of the characters are processed further in methods called from the
current method, we replace the portions of expansion that is processed by nested
method calls by the \nonterminal symbol corresponding to the nested method call.
When, say, \texttt{parse\_expr()} uses the result of \texttt{parse\_num()}, this
result will be referred to as the \texttt{parse\_num} \nonterminal.

\item Methods often have control structures such as if-conditions that
selectively process parts of the input. Further, methods may also contain
looping structures that process repetitive parts of the input.
To handle these cases, we adapt
\emph{regular right hand sides}~\cite{jim2010efficient}
for the grammar\footnote{A grammar with regular right hand sides is different
from a typical context-free grammar in that it allows regular expressions in the
rules. EBNF is a notable example.}.
Input characters processed inside any if/else conditions are turned into
\emph{regular expression alternations}. Finally, inputs processed inside
loops, after recovering the repetition order through active learning\footnote{
The term \emph{active learning} was first used by Dana
Angluin~\cite{angluin1987learning} in the context of regular grammar learning.
},
are turned into regular expression groups with \emph{Kleene star}.
If \texttt{parse\_num()} uses a loop to read in digits, we will identify
that \texttt{parse\_num} consists of repeated digits, resulting in the
regular expression \texttt{[0-9]+}.

\item Each method call becomes a named node with the method name. Each
iteration becomes a node in the parse tree with the name derived from the
method name, loop name and the location of the loop starting. Finally, each
conditional becomes a node in the parse tree.
\end{enumerate}

As an example, consider the recursive descent parser in \Cref{fig:calcprogram}.
Running it with an argument
\texttt{9+3/4} yields the tentative parse tree shown in Figure~\ref{fig:derivationtree}.
We extract such parse trees for a number of given inputs. Next, we traverse each tree and identify \emph{loop} nodes that are similar as we detail in Section~\ref{sec:activelearning}. This results in parse trees where similar nodes have similar names. Finally, we construct the grammar by recursively
traversing each parse tree, collecting the name and children types and names for each node. The node names become \nonterminal symbols in the grammar, and each set of children becomes one possible expansion in the grammar being constructed. The child nodes that represent characters become terminal symbols in the constructed grammar.

The final result is the grammar in \Cref{fig:calcgrammar}, which exactly reflects the capabilities of the program in \Cref{fig:calcprogram}.  Again, the grammar is readable with near-textbook quality and well reflects the input structure.  Using this grammar as a producer yields an arbitrary large number of syntactically valid input strings.
%
\begin{comment}
%
that quickly cover all code.\footnote{Except for the error-handling parts; to test these, a random character generator would do well.}
\end{comment}

%
%
%
%

In the remainder of this paper, we detail our contributions:
\begin{enumerate}
  \item We provide a general algorithm for deriving the context-free
    approximation of input language from a recursive descent parser.
    Our approach relies on \emph{tracking character access in the input buffer}
    (\Cref{sec:instrumentation}), which is easy to implement for a variety of
    languages that support string wrappers, or the source can be transformed to
    support such wrappers. From the tracked accesses, we then infer
    \emph{parse trees} (\Cref{sec:parsetree}), which we generalize by means of
    \emph{active learning} before passing them to our grammar inference
    (\Cref{sec:inference}).  Our approach leverages structure of input
    processing and its identifiers to produce input grammars that are fit
    for human consumption out of the box.

  \item We evaluate our approach, comparing it against the so far only approach for inference of human-readable grammars (\AUTOGRAM).  For the evaluation, we use producers (\Cref{sec:producer}) to assess recall and precision (\Cref{sec:evaluation}).  Our approach is superior to \AUTOGRAM in both aspects.

  \item In our evaluation, we also show that our approach is applicable in
    contexts in which no usable data flow of input fragments to variables
    exists, as well as for advanced parsers
    such as \emph{PEG} parsers and \emph{parser combinators} which make
    the state of the art for writing secure parsers~\cite{chifflier2017writing}.
    None of these is possible with \AUTOGRAM, again extending the state of the art.
\end{enumerate}

After discussing limitations (\Cref{sec:limitations}) and related work (\Cref{sec:background}), \Cref{sec:conclusion} closes with conclusion and future work.  The complete source code of our approach and evaluation is available.

\iflong
The remainder of the paper is organized as follows.  We start by describing the individual steps of our technique. \Cref{sec:instrumentation} details how we track the control flow and character accesses. \Cref{sec:parsetree} shows how to turn the resulting traces into parse trees.  \Cref{sec:inference} details
the algorithm for inferring the grammar from the parse trees.  \Cref{sec:producer} finally sketches how to use these grammars for producing valid inputs.  All this is then put to test in our evaluation (\Cref{sec:evaluation}), pitching our \mimid tool against the \AUTOGRAM state of the art.  After discussing limitations (\Cref{sec:limitations}) and the state of the art (\Cref{sec:background}), we close with conclusion and future work in \Cref{sec:conclusion}.
\fi

\section{Tracking Control Flow and Comparisons}
\label{sec:instrumentation}

Let us now start with describing the details of how we infer input grammars from
dynamic control flow.  We start with the \emph{tracing} part---that is, acquiring dynamic control flow information from a program run.

For our experiments, we used the Python language, as Python makes it fairly easy to implement dynamic    analysis techniques.  Other than dynamic analysis, we do not make use of specific Python features, and    implementing the techniques for other languages should be feasible with modest effort.

For tracking the control flow, we programmatically modify the parser source. We
insert a tracker for both method entry and exit as well as trackers for control
flow entry and exit for any conditions and loops. For the purposes of our
approach, we consider these control structures as \emph{pseudo methods}.
Every such method call (both true and pseudo method calls) gets a unique
identifier from a counter such that a child of the current method or a later
method call gets a larger new method identifier than the current one.
We note that this information --- the execution tree --- can be cheaply
obtained from programs with simple coverage instrumentation.

For tracking the character accesses being made, we simply wrapped the input
string in a proxy object that logged access to characters. We annotate each
character accessed with the current method name.

During our analysis, we also check whether parser code guarded
by conditionals can be skipped completely. That is, for cascading conditionals,
we check whether the conditional ends in an \emph{else} condition or not.
Each method is also annotated with the current stack of pseudo methods until
the most recent parent true method. Thus, each character access is
associated with a specific method call, and each method call contains the
information about the current stack of pseudo methods.

\section{From Traces to Parse Trees}
\label{sec:parsetree}

The instrumented program is provided with a \emph{sample input} to collect the
\emph{character access} traces on associated control-flow nodes.
Once we have the \emph{character accesses} on the indexes on the origin
string, we have to decide which method call should be associated with the
particular character index.
We follow a simple strategy:

\textbf{The \emph{last method} to access a particular character index consumes it.}

Next, we need to eliminate overlaps between child nodes of the same node or
between the parent and the child. That is, we want each child to own a single
contiguous index range.
To do that, we recursively scan the tree, and verify that the begin and end
of none of the children of the same node overlap. If an overlap is found, the
tie is decided in favor of the child that accessed the part last. The child that
is in the overlap is recursively scanned, and any children that are contained in
the overlap are removed.

\begin{figure*}
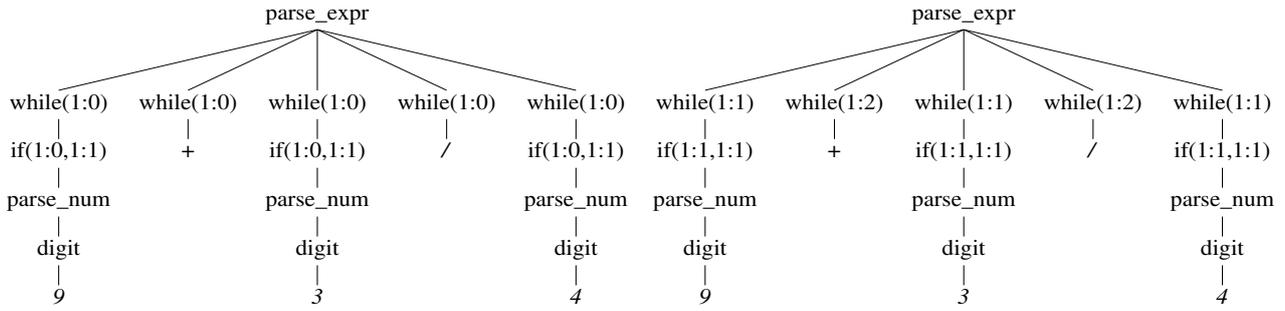

\begin{subfigure}{.49\textwidth}
   \resizebox{.99\columnwidth}{.05\totalheight}{
\Tree[.parse\_expr
  [.while(1:0)
                      [.if(1:0,1:1)
                        [.parse\_num
                          [.digit
                        \textit{9}
                          ] ] ] ]
  [.while(1:0) \textit{+} ]
  [.while(1:0)
                      [.if(1:0,1:1)
                        [.parse\_num
                          [.digit
                        \textit{3}
                          ] ] ] ]
  [.while(1:0) \textit{/} ]
  [.while(1:0)
                      [.if(1:0,1:1)
                        [.parse\_num
                          [.digit
                        \textit{4}
                          ] ] ] ]
                          ]}
\caption{Non-generalized parse tree.
  The number before colon indicates the identifier of the particular
  control flow pseudo-method, and after number after the colon identifies the
  alternative if any. That is, \texttt{if(... 5:2)} indicates the fifth
  \texttt{if} whose \texttt{else} branch was taken.
  The pseudo method stack is in the parenthesis.}
\label{fig:ngderivationtree}
\end{subfigure}
\begin{subfigure}{.10\textwidth}
\end{subfigure}
\begin{subfigure}{.49\textwidth}
  \resizebox{.99\columnwidth}{.05\totalheight}{
\Tree[.parse\_expr
 [.while(1:1)
                     [.if(1:1,1:1)
                       [.parse\_num
                         [.digit
                       \textit{9}
                         ] ] ] ]
 [.while(1:2) \textit{+} ]
 [.while(1:1)
                     [.if(1:1,1:1)
                       [.parse\_num
                         [.digit
                       \textit{3}
                         ] ] ] ]
 [.while(1:2) \textit{/} ]
 [.while(1:1)
                     [.if(1:1,1:1)
                       [.parse\_num
                         [.digit
                       \textit{4}
                         ] ] ] ]
                         ]}
\caption{Generalized parse tree. The
 number in suffix after colon indicates the generalized identifier after validating replacements.
 As before, the pseudo method stack is contained in the parenthesis,
 which is also updated when the parent is updated during generalization.}
\label{fig:gderivationtree}
\end{subfigure}
\caption{Parse trees for \texttt{9+3/4}.The prefix before colon indicates the static identifier of the control
structure in the method. That is, the first \sif gets the prefix \texttt{1:}. The
suffix is explained above.}
\end{figure*}

Once the indexes are associated with method call identifiers, we generate
a \emph{call tree} with each method identifier arranged such that methods called
from a given method are its children. The last accessed input
indexes are added as the leaf nodes of the tree. As there is no overlap,
such a tree can be considered as a \emph{parse tree} for the given input string.
The parse tree at this point is given in Figure~\ref{fig:ngderivationtree},
which we call the \emph{non-generalized} parse tree of the input string.
In Figure~\ref{fig:ngderivationtree}, each pseudo method has a list of values
in parenthesis, in the following format. The last value in the parenthesis is
the identifier for the control flow node taken. That is given a node name as
\texttt{if(2:0, 3:1)}, the identifier is \texttt{3:1}. It indicates that the
corresponding \texttt{if} statement was the third conditional in the program,
and the execution took the first (\emph{if}) branch of the conditional.
If the identifier was \texttt{3:2}, it would indicate that the execution took
the \emph{else} branch, and for larger values, it indicates the corresponding
branch of a cascading if statement or a case statement. In the case of loops,
there is only a single branch that has child nodes, and hence is this is
indicated by \texttt{0}. The values before the identifier corresponds to the
identifiers of the pseudo-method parents of this node until the first method
call. That is, the \texttt{if(2:0, 3:1)} has a parent node that is a loop, and
it was the second loop in the program.

While we have produced a parse tree, it is not yet in a format from which we
can recover the context-free grammar. To be able to do so, we need accurately
labeled parse trees where any given node can be replaced by a node of similar
kind without affecting the (parse) validity of the string.
The problem here is that, not all iterations of loops are replaceable with each
other. That is, loops can be influenced by the previous iterations. For example,
consider the derivation tree in Figure~\ref{fig:ngderivationtree}.
If one considers each iteration of the loop to be one \emph{alternate expansion}
to the corresponding \nonterminal, the rule recovered is:
\begin{grammar}\centering
  <expr> $\rightarrow$ \rhs{num | + | /}
\end{grammar}
However, this is incorrect as a single free-standing operator such as
\texttt{+} is not a valid value. The problem is that \texttt{is_op} encodes
a link between different iterations. Hence, we annotate each individual
iteration, and leave recovering the actual repeating structure for the next
step. A similar problem occurs with method calls too. In the parse tree we
produced, we assume that any given \nonterminal --- say \texttt{parse\_num}
can be replaced by another instance of \texttt{parse\_num} without affecting
the validity of the string. However, this assumption may not hold true in
every case. The behavior of a method may be influenced by a number of factors
including its parameters and the global environment. We fix this in the
next step.

\subsection{Active Learning of Labeling}
\label{sec:activelearning}
The parse tree thus derived is accurate but not very useful to construct a
grammar. The problem is the unique labeling of nodes. We do not
yet know the dependencies between each iteration. Nor do we know if the behavior
of any method is influenced by its parameters or its environment. To determine
the precise labeling, we simply traverse all parse trees we have, and collect
every single node, and separate them by the name of the node. That is, all
\texttt{parse\_num} nodes go together, so does all \texttt{if(1:0, 1:1)}.

We now want to identify whether each node that is grouped under a node name is
replaceable (or compatible) with another with the same name. Unfortunately,
compatibility is not \emph{transitive} if one is looking at parse validity.
For example, say, there are three \emph{words} in a language --- \texttt{a},
\texttt{b}, and \texttt{ac}. Each \emph{word} is composed of individual
\emph{letters}. In the case of \texttt{a}, and \texttt{b}, the corresponding
letter, and for \texttt{ac}, the letters \texttt{a}, and \texttt{c}.
\begin{grammar}
<START> ::= <word1> | <word2> | <word3>

<word1>  ::= <letter\_a>

<word2>  ::= <letter\_b>

<word3>  ::= <letter\_a><letter\_c>

<letter\_a> ::= `a'

<letter\_b> ::= `b'

<letter\_c> ::= `c'
\end{grammar}
Now, consider the parse trees of \texttt{a}, \texttt{b}, and \texttt{ac}.
\begin{lstlisting}
 (START (word1 (letter_a "a")))
 (START (word2 (letter_b "b")))
 (START (word3 (letter_a "a") (letter_c "c")))
\end{lstlisting}
We consider a node as compatible with another if the string produced from
a parse tree where the first node is replaced by the second is parsed correctly
-- that is, the generated string parses without any errors, and the parse
tree generated from the new parse has the \emph{same structure} as the tree
generated by replacing the node.

Here, the nodes \texttt{letter\_a} across parse trees are compatible because
the generated strings are exactly the same. Next, the \texttt{letter\_a} under
\texttt{word1} is compatible with \texttt{letter\_b} under
\texttt{word2}. The generated strings are \texttt{a} and \texttt{b}. So,
is the node \texttt{letter\_b} under \texttt{word2} compatible with
\texttt{letter\_a} under \texttt{word3}? Unfortunately not, as the string
generated from
\begin{lstlisting}
  (START (word3 (letter_b b) (letter_c c)))
\end{lstlisting}
is \texttt{bc} which is not in the language.

This means that for accurate identification of compatible nodes, each node
has to be compared with all other nodes with the same name, which gives us
a complexity of $O(n^2)$ in the worst case in terms of the number of nodes.
However, we found that the assumption of \emph{transitivity} rarely breaks,
and even then, the inaccuracy induced, affects less than $10\%$ of inputs
generated from the grammar (See the evaluation of \mathexpr).
Since the assumption of \emph{transitivity}
allows us to reduce the computational effort, our evaluation is implemented
assuming \emph{transitivity} of compatibility.\footnote{
We note that one does not need to rely on this assumption. One can choose
to do the complete $O(n^2)$ verification, or can choose anything in between
that and the faster but approximate version.
}

Once we have identified the compatibility buckets, we can update the nodes
in them with unique suffixes corresponding to each bucket, and update the
node name of each with the suffix.
In the case of loop nodes, we also update the stack name of this node in all the
child and grand child elements of this node --- all grand children up to the
next non-pseudo method call. The idea here is that, if there are two unique loop
iterations that are incompatible with each other, then any other control flow
nodes inside that loops such as conditionals should also be considered
incompatible even if the same alternative path is taken in the conditional
during the execution of both iterations.

Once the process is complete, all the nodes in all the parse trees will be
labeled with consistent and correct identifiers. These can then be extracted to
produce the correct grammar. The generalized counterpart to Figure~\ref{fig:derivationtree}
is given in Figure~\ref{fig:gderivationtree}.

\subsection{Active Learning of Nullability}

For conditional nodes, whether an \texttt{if} node can be skipped can
be determined statically without active learning, by simply checking for
the presence of an \texttt{else} branch.

However, unlike conditionals, we do not have a simple way to statically
determine if a loop can be skipped entirely or at least one iteration is
required. One alternative is to wait for sufficient number of samples, and
see if there are examples where the loops are absent. A second alternative
is to use \emph{active learning}.

The idea is to replace all \emph{consecutive} \sloop nodes that are
the children of a given node in a given parse tree. Then check the validity
of the string produced from that tree. If the parse structure of the
new string is correct, and this can be done on all parse trees and at all
points where this is possible, the loop is marked as nullable.

Similar to loops, \sif conditionals may also be labeled incorrectly. For
example, consider the set of statements below.
\begin{lstlisting}[style=python]
if g_validate:
  validate_header(header)
\end{lstlisting}
The problem here is that, while the \sif does not have an \selse branch,
we do now know whether the body of the conditional can be skipped or not.
In particular, the \texttt{g\_validate} may be a global configuration option
which may mean that it is always enabled or always disabled for specific
kinds of parse trees. While we have not found such conditionals in our
subjects, if additional accuracy is desired, the optional parts of conditionals
may also be verified using active learning.

With this, our trees are accurately labeled and ready for inferring grammars from them.

\section{Grammar Inference}
\label{sec:inference}

The basic idea of constructing a grammar out of a labeled parse tree is simple.
We traverse each parse tree starting from the top, descending into each
children, and each node we see, if it is not a character node, is marked as
a \nonterminal in the grammar. The children are placed as the rule for expansion
of the \nonterminal in the grammar. If the child is a non-character node, the
token in the expansion will be a reference to the corresponding \nonterminal
in the grammar. There may be multiple alternate expansions to the same
\nonterminal even from the same tree as the same method call may be made
recursively. This is detailed in Algorithm~\ref{alg:basicgrammar}.
\begin{algorithm}[H]
 \begin{algorithmic}
 \Function{extract\_grammar}{node, grammar}
  \State $\textrm{name}, \textrm{uid}, \textrm{children}, \textrm{stack} \Leftarrow \textrm{node}$
  \State $\textrm{a_name} \Leftarrow \textrm{name} + \textrm{uid} + \textrm{stack}$
  \State $\textrm{rule} \Leftarrow []$
  \If {$\textrm{a_name} \not\in grammar$}
  \State $\textrm{grammar}[\textrm{a_name}] \Leftarrow \{\textrm{rule}\}$
   \Else
  \State $\textrm{grammar}[\textrm{a_name}].add(\textrm{rule})$
   \EndIf
  \If {$\textrm{children} = \emptyset$}
    \State \textbf{return} $\textrm{T},\textrm{a_name}$
   \Else
   \For{$\textrm{child} \leftarrow \textrm{children}$}
     \State $\textrm{kind}, \textrm{cname} \Leftarrow \textrm{extract\_grammar}(\textrm{child})$
    \If {$\textrm{kind} = \textrm{T}$}
    \State $\textrm{rule } += \textrm{to\_terminal}(\textrm{cname})$
     \Else
    \State $\textrm{rule } += \textrm{to\_nonterminal}(\textrm{cname})$
     \EndIf
   \EndFor
   \EndIf
  \State \textbf{return} $\textrm{NT}, \textrm{a_name}$
 \EndFunction
 \end{algorithmic}
   \caption{Extracting the basic grammar}
  \label{alg:basicgrammar}
\end{algorithm}

\iflong
A complication in grammar inference is that our rules may look
as below (method stack details are skipped).
\settowidth{\widest}{aaaaaaaaaaaaaaaaaaaaaaaaaaaaaaaaaaaaa}
\begin{grammar}
  <expr>  $\rightarrow$ \rhs{while:1}

  <expr>  $\rightarrow$ \rhs{while:1 while:2 while:1 while:2 while:1}

  <expr>  $\rightarrow$ \rhs{while:1 while:2 while:1}

  <while:1>  $\rightarrow$ \rhs{if:1}

  <if:1>  $\rightarrow$ \rhs{num}

  <while:2>  $\rightarrow$ \rhs{+}
\end{grammar}
There may be even more complex repeating patterns. The problem is that we want
to merge and abstract these rules because they represent possible infinite
repetitions. We see next how they can be handled.
\fi

\iflong
\subsection{Identifying Repeating Patterns}
\fi

An additional challenge comes from identifying repeating patterns.
\iflong
We essentially want
to identify the repeating patterns even if they are a few levels deep, and
we want to identify the best repeating patterns here.
\fi
Fortunately, this
problem has various solutions~\cite{fernau2009algorithms}. We chose a
modification of the \emph{prefix tree acceptor} algorithm\footnote{
Unlike the original \pta, which considers only repeating patterns single
character long, we first scan for, and identify repeating patterns
of any block size. We next scan the inputs for any instances of the identified
repeating patterns. These are then chunked, and considered as the alphabets
as in the original \pta algorithm.
}.
Once we run the modified \pta algorithm, the previous grammar is
transformed to:
\begin{grammar}
  <expr>  $\rightarrow$ \rhs{while:1}

  <expr>  $\rightarrow$ \rhs{(while:1 while:2)+ while:1}

  <while:1>  $\rightarrow$ \rhs{if:1}

  <if:1>  $\rightarrow$ \rhs{num}

  <while:2>  $\rightarrow$ \rhs{+}
\end{grammar}
The \texttt{while:<n>} can be replaced by the regular expression summaries
of the corresponding rules recursively. Here, this gives us the regular
expression:
\begin{grammar}
  <expr>  $\rightarrow$ \rhs{num | (num [+])+ num}
\end{grammar}
While generalizing, we can replace any $+$ with $*$ provided all the items
inside the group are nullable. Similarly, when merging regular expressions
corresponding to conditionals, one can add $(...|)$ i.e. an $\epsilon$
alternative, provided the corresponding if condition was nullable.
These steps generate the right hand side regular expressions in \Cref{fig:calcgrammar}
for a simple program given in \Cref{fig:calcprogram}. For details on learning
regular expressions from samples, see Higuera~\cite{higuera2010grammatical}.
The grammar derived from \microjson after removing differences due to
\emph{white space} is given in \Cref{fig:mimidjson}.

\subsection{Producing a Compact Grammar}
At this point, the mined grammar readable but verbose. There are a number of
transformations that one can take to reduce its verbosity without changing
the language defined by the grammar. These are as follows:

\begin{enumerate}
  \item If there is any key that is defined by a single rule with a single
    token, delete the key from the grammar, and replace all references to that
    key with the token instead.
  \item If there are multiple keys with the same rule set, choose one, delete
    the rest, and update the references to other keys with the chosen one.
  \item If there are duplicate rules under the same key, remove the redundant
    rules.
  \item Remove any rule that is same as key it defines.
  \item If there is any key that is referred to on a single rule on a single
    token, and the key is defined by just one rule, delete the key from the
    grammar, and replace the reference to the key by the rule.
\end{enumerate}

We repeat these steps as long as the number of rules in the grammar decreases
in each cycle. This produces a smaller grammar that defines the same language.

\section{Generating Inputs from Grammars}
\label{sec:producer}

\begin{figure}[H]\footnotesize
\def\sep{\quad\cdot\quad}
$
5 \sep
(8+(2-((338+50409)/56))) \sep
((8/1))-7-8-7-((9)+9-3/25) \sep
5/((6-(88233/(60)))) \sep
4*(8-((99/((1+3)))-31200/(8+(308)))) \sep
(((2/0)))+(39+(4*2*70)) \sep
((9*(1*3))*((0*(0))/((5)-(9+8))))+(2) \sep
(802+3)/(1+(5758*(74506+(((77))+902369)+2))) \sep
(854*1+(3)-((37*((2))/7-(((8-(((((084/((((((9/(5))/(((((0-((((9))+97966)-(1-81)))*33)+(((982+7)-20932)-66))-659))/2)+26)-681)-519+76)+225))/489))/15)))*((9-(0))-(7)))*08))))
$
\caption{Inputs produced from the grammar in \Cref{fig:calcgrammar}}
\label{fig:calcfuzz}
\end{figure}

Once a grammar is extracted, it can immediately be turned it into a \emph{producer}, which starting from the start symbol, will apply one expansion after the other to produce inputs.  For our calculator example from \Cref{fig:calcprogram}, the extracted grammar in \Cref{fig:calcgrammar} yields arithmetic expressions such as the ones shown in \Cref{fig:calcfuzz}.
State of the art tools like \emph{F1}~\cite{gopinath2019building} implement several optimizations to prevent out-of-bound growth.  In our experiments, we make use of the Python \emph{GrammarCoverageFuzzer}~\cite{fuzzingbook2019:GrammarCoverageFuzzer}, which additionally aims for systematically covering all input elements.

\iflong It does not take a grammar for fuzzing, though.  As our seed inputs are all decomposed into      parse trees, one can also mutate and recombine these parse trees to obtain large sets of test             samples~\cite{holler2012langfuzz}.  This alternative is especially valuable when applying our approach on parsers that do not produce named \nonterminals (e.g. PEG parsers and parser combinators).\fi

\section{Evaluation}
\label{sec:evaluation}

For evaluation, we used the following subjects:
\begin{description}
  \item[Calculator] (\calc) -- a simple recursive descent program written
    in textbook style
    from
    the Codeproject\footnote{\url{https://www.codeproject.com/Articles/88435/Simple-Guide-to-Mathematical-Expression-Parsing}},
    simplified and converted to Python.
    It also forms the running example in the paper.
    We used self generated expressions to mine the grammar and evaluate.
  \item[Mathexpr] (\mathexpr) -- a more advanced expression evaluator from the
   Codeproject%
   \footnote{\url{https://github.com/louisfisch/mathematical-expressions-parser}
  }.
   It includes pre-defined constants, method calls, and the ability to define variables.
   As in the case with Calculator, we used self generated expressions
    and test cases to mine the grammar and evaluate.

  \item[CGIDecode] (\cgidecode) -- the \emph{CGIDecode} program originally from the Pezze et al.~\cite{pezze2008software},
    with the Python implementation from the chapter on \emph{Code Coverage}~\cite{fuzzingbook2019:Coverage} from Zeller et al.
  This is an example of a parser that is a simple state machine. It is not recursive, and hence does not
  use the stack. For \emph{CGIDecode}, we used self generated expressions to mine the grammar and evaluate.

  \item[URLParse] (\urlparse) -- the URL parser part of the Python \emph{urllib}
 library\footnote{\url{https://github.com/python/cpython/blob/3.6/Lib/urllib/parse.py}}. An example of
  ad hoc parsing with little ordering between how the parts are parsed.
  For initial mining and evaluation, we used the URLs generated from passing tests using the \emph{test\_urllib.py} in
  the Python distribution. We also used a handwritten grammar to generate inputs as we detail later.

  \item[NetRC] (\netrc) -- the \emph{netrc} library from the Python distribution\footnote{\url{https://  github.com/python/cpython/blob/3.6/Lib/netrc.py}}
    which is used to parse the \emph{.netrc} files that contain login information.
  This parser uses a separate lexing stage with the \emph{shlex} lexer from the Python
  distribution. The \emph{lexing} stage can stop the grammar recovery completely
  because we stop tracking the character access as soon as it is transformed from
  a character stream to a token. Hence, for NetRC, we modified the \emph{shlex} to transmit proxy
  strings that allow one to track character accesses similar to object
  based tainting~\cite{conti2010taint,fuzzingbook2019:InformationFlow}.
  For \emph{NetRC}, few samples were available in \emph{test_netrc.py}. Hence,
  we searched for samples of \emph{.netrc} online and found ten samples. These
  were used as a template for writing a grammar
  for the \emph{NetRC} format, which was used to generate inputs.

  \item[MicroJSON] (\microjson) -- a minimal JSON parser from Github\footnote{\url{https://github.com/phensley/microjson}}.
  We fixed a few bugs in this project during the course of extracting its grammar (merged upstream).
  For mining, we chose ten simple samples that explored all codepaths in the parser.
  For our evaluation, we used 100 samples of JSON generated from the following
  JSON API end points: \emph{api.duckduckgo.com},	\emph{developer.github.com},
  \emph{api.github.com}, \emph{dictionaryapi.com}, \emph{wordsapi.com}, \emph{tech.yandex.com}.
  We also added sample JSON files obtained from \emph{json.org}, \emph{json-schema.org},
  \emph{jsonlint}, \emph{www.w3schools.com}, and \emph{opensource.adobe.com}.
\end{description}

Are our subjects too few or small? We note that our focus is on recovering
input grammar from parsers written in various recursive descent parsing styles.
It would be obvious to parser implementers that there is little difference
between a small and a large parser in terms of its implementation other than the
scale. To buttress the point, if one has an off the shell parser, there is
little difference between a large grammar or a small grammar. Similarly, for a
particular style of parsing, showing that one can recover the grammar from a
small program is no different from showing that one can recover the input
grammar from a larger one. Hence, our focus was to identify parsers that deploy
diverse strategies for parsing (while still being recursive descent).

Indeed, the subjects for our evaluation have diverse strategies for parsing. For
example, the \cgidecode is an automata, and recognizes a regular language.
\urlparse also recognizes a regular language, but is written in an ad hoc style
with parts being parsed in a different order than it appears in the input.
The \calc is an example of an text book style procedure based approach to
parsing, and uses an input buffer and an index to specify the current parsing
location. It recognizes a context free language for arithmetic expressions.
The \mathexpr is
more complex. It has an object oriented structure, and the input being parsed is
encapsulated within the object. It recognizes a context free language for
arithmetic expressions with variables and functions. \microjson recognizes JSON, which is a
real world context-free format, and contains complex parsing rules. Finally,
\netrc also recognizes a context-free language, but in addition, sports a
\emph{lexing} stage.

\subsection{Methodology}

For comparing our approach with \AUTOGRAM, we used the Python implementation
in the chapter on \emph{Mining Input Grammars}~\cite{fuzzingbook2019:GrammarMiner}
from Zeller et al.  \emph{GrammarMiner} is a Python implementation of \AUTOGRAM with
important differences. The first is that, unlike the original approach,
the \emph{GrammarMiner} is written in Python and is used for analysis of Python programs.
It is a technology demonstrator used for illustrating the concept of
grammar mining. Hence, our comparisons may not apply directly to the original
\AUTOGRAM implementation. However, our focus is on the \emph{conceptual}
differences between \AUTOGRAM and \Mimid, especially whether there are programs
that can not be easily analyzed by the \AUTOGRAM approach of using data flow
to recover the program input grammar. For all differences in results, we thus investigate how much they are due to \emph{conceptual} differences.

We note that the Python implementation of \AUTOGRAM does not implement
generalization of character sets to regular expressions unlike the original
\AUTOGRAM. For a fair comparison, we have disabled \Mimid generalization
of character sets to larger regular expressions during comparisons.

Our research questions are as follows:
\begin{description}
  \item[RQ 1.] How accurate are the derived grammars as \emph{producers?}
  \item[RQ 2.] How accurate are the derived grammars as \emph{parsers?}
  \item[RQ 3.] Can one apply \Mimid to parsers without data flow?
  \item[RQ 4.] Can one apply \Mimid to non-traditional parsers such as PEG and Parser combinators?
\end{description}

\subsection{RQ 1. Grammar Accuracy as Producers}

As we explained previously, we collected a set of detailed samples for
each of the programs for grammar mining. Next, we used both \AUTOGRAM and \Mimid on the
\emph{same} set of samples and generated a grammar from each approach for each
program. This grammar was then used to generate inputs to fuzz the program using
the \emph{GrammarFuzzer}~\cite{fuzzingbook2019:GrammarFuzzer}.
The number of inputs that were accepted by the subject program is given
in Table~\ref{tbl:fuzz}.

\begin{table}[H]
  \centering
\caption{Inputs generated by inferred grammars that were accepted by the program (1,000 inputs   each)}
\rowcolors{1}{Goldenrod}{white}
\begin{tabular}{rrr}
          & \AUTOGRAM & \Mimid \\
  \hline
  \calc & 36.5\% & 100.0\% \\
  \mathexpr & 30.3\% & 87.5\% \\
  \cgidecode & 47.8\% & 100.0\% \\
  \urlparse & 100.0\% & 100.0\% \\
  \netrc & 0.0\% & 100.0\% \\
  \microjson & 53.8\% & 98.7\% \\
\end{tabular}
\label{tbl:fuzz}
\end{table}

\begin{result}
Grammars inferred by \Mimid \emph{produce} many more correct inputs than \AUTOGRAM.
\end{result}

To assess and understand the differences in results, we manually examined the grammars from both         \AUTOGRAM and \Mimid.

\subsubsection{\textbf{\calc}}

A snippet of the grammar from \AUTOGRAM for \calc is below.
\begin{grammar}
<START> ::= <init@884:self>

<init@884:self> ::= <expr@26:c>00
\alt <expr@26:c>3<expr@26:c><expr@29:num>\\
  <expr@26:c>*<expr@29:num>*4
\alt <expr@26:c>1<expr@26:c><expr@26:c>0<expr@26:c>2
\alt <expr@26:c>100)
...
\end{grammar}
The grammar from our approach for the same program, using same mining sample
is given in Figure~\ref{fig:calcgrammar}. An examination shows that the rules
derived by \AUTOGRAM were not as general as \Mimid. That is, the grammar
generated by \AUTOGRAM is enumerative at the expense of generality. Why
does this happen? The reason is that the \texttt{parse\_expr} and other
functions in \calc accept a \emph{buffer} of input characters, with an index
specifying the current parse location. \AUTOGRAM relies on
fragmented values being passed in to method calls for successful extraction
of parts, and hence the derivation of tree structure. Here, the first call
creates a linear tree with each nested method claiming the entirety of the
buffer, and this defeats the \AUTOGRAM algorithm.

This is not a matter of better implementation. The original \AUTOGRAM
relies on parameters to the method calls to contain only parts of the input.
While an \AUTOGRAM derived algorithm may choose to ignore the method parameters,
any use of a similar buffer inside the method will cause the algorithm to fail
unless it is able to identify such spurious variables.

\subsubsection{\textbf{\mathexpr}}

For \mathexpr, the situation was again similar. \AUTOGRAM was unable to
abstract any rules. The \mathexpr program uses a variation of the strategy
used by \calc for parsing. It stores the input in an internal buffer in the
class and stores the index to the buffer as the location being currently
parsed. For similar reasons as before, \AUTOGRAM fails to extract the parts
of the buffer. \Mimid on the other hand produced an almost correct grammar,
correctly identifying constants and external variables. The single mistake
found (which was cause of multiple invalid outputs) was instructive.
The \mathexpr program pre-defines letters from
\texttt{a} to \texttt{z} as constants. Further, it also defines functions
such as \texttt{exp()}. The function names are checked in the same place
as the constants are parsed. \Mimid found that the function names are
composed of \emph{letters}, and some of the letters in the function names
are compatible with the single letter variables --- in that, they can be
exchanged and still produce correct values. Since we assumed
\emph{transitivity}, \Mimid assumed that all letters in function names
are compatible with single letter constants. This assumption produced
function names such as \emph{eep()}, which failed the input validation.

\subsubsection{\textbf{\microjson}}

The inferred \microjson with \AUTOGRAM produces more than 50\%, valid
inputs when compared to 98.2\% from \Mimid. Further, we note that
the 50\% valid inputs from \AUTOGRAM paints a more robust picture than
the actual situation. That is, the grammar recovered by \AUTOGRAM for
\microjson is mostly an enumeration of the values seen during mining.
The reason is that \microjson uses a data structure \texttt{JStream} which
internally contains a \texttt{StringIO} buffer of data. This data structure
is passed around as method parameters for all method calls, e.g.
\texttt{\_from\_json\_string(stm)}. Hence, every call to the data structure
gets the complete buffer with no chance of breaking it apart. We note that
it is possible to work around this problem by essentially ignoring method
parameters and focusing more on return values. The problem with the data
structure can also be worked around by modifying the data structure to hold
only the remaining data to be parsed. This however, requires some specific
knowledge of the program being analyzed. \Mimid on the other hand is not
affected by the problem of buffers at all and recovers a complete grammar.

\subsubsection{\textbf{\urlparse}}

For \urlparse we initially noticed that \AUTOGRAM and \mimid did not perform
well (the inferred grammar could recognize less than 10\% of the samples)
due to the inability to generalize strings.
Since we were interested in comparing the capabilities
of both algorithms in detecting structure, we restricted our mining sample
to only contain a specific set of strings. In particular, we noticed that
the \urlparse program split the given input to
\texttt{<scheme>}, \texttt{<netloc>}, \texttt{<query>}, and
\texttt{<fragment>} based on delimiters. In particular, the internal
structure of \texttt{<netloc>} and \texttt{<query>} were ignored.

Hence, we wrote a grammar for \urlparse which contained a list of specific
strings for each part. Next, we generated 100
inputs each using the Grammar Fuzzer and validated each by checking the
string with the program. We then used these strings as mining set. With
this mining set, the grammar from both \Mimid and \AUTOGRAM could produce
100\% correct inputs.

\subsubsection{\textbf{\netrc}}
For \netrc, similar to \urlparse we noticed that \AUTOGRAM and \mimid did
not perform well due to the inability to generalize strings.
Since we were interested in comparing the capabilities
of both algorithms in detecting structure, we restricted our mining sample
to only contain a specific set of strings including fixed spaces.
Next, we generated 100 inputs each using the Grammar Fuzzer and validated each
by checking the string with the program. We then used these strings as mining
set. With this mining set, the grammar mined using \Mimid was able to recognize
100\% of the input. Unfortunately, \AUTOGRAM was unable to produce valid
inputs. The problem is that, \netrc employs \texttt{shlex}, a lexical stage
before parsing. The \emph{lexing} stage seems to have confused \AUTOGRAM, especially
for the first token, where all characters in the token are assigned to the same
\nonterminal symbol. This essentially leads to incorrect inputs even though
only the first token is incorrect in a large number of inputs.

If one fixes this by hand in the inputs, \AUTOGRAM produces 69.2\% valid
inputs, which was still lesser than \mimid.

\subsection{RQ 2. Grammar Accuracy as Parsers}

For our second question, we want to assess whether correct inputs would also be accepted by our inferred grammars.  In order to obtain correct inputs, we used various approaches as available for different
grammars. For \calc and \mathexpr, we wrote a grammar by hand.
Next, we used this grammar to generate a set of inputs
that were then run through the subject programs to check whether the inputs were
valid. We collected 1,000 such inputs for both programs. Next, these inputs were
fed into parsers using grammars produced by \Mimid and \AUTOGRAM.
We used the \emph{Iterative Earley Parser} from~\cite{fuzzingbook2019:Parser}
for verifying that the inputs were parsed by the given grammar.

For \netrc and \urlparse, we used the same grammar for parsing that we already had used to generate
mining inputs. We again collected a set of valid inputs and verified that
the inferred grammar is able to parse these inputs.
For \microjson, we used the collected JSON documents as described above.
The largest document was 2,840 lines long. We then verified whether the grammar inferred
by each algorithm was able to parse these inputs. Our results are given in Table~\ref{tbl:recall}.
\begin{table}[H]
  \centering
\caption{Inputs generated by golden grammar that were accepted by the
  inferred grammar parser (1,000 inputs each except \microjson which used 100 external inputs)}
\rowcolors{1}{Apricot}{white}
\begin{tabular}{rrr}
          & \AUTOGRAM & \Mimid \\
  \hline
  \calc & 0.0\% & 100.0\% \\
  \mathexpr & 0.0\% & 92.7\% \\
  \cgidecode & 35.1\% & 100.0\% \\
  \urlparse & 100.0\% & 96.4\% \\
  \netrc & 34.6\% & 41.3\% \\
  \microjson & 0.0\% & 93.0\% \\
\end{tabular}
\label{tbl:recall}
\end{table}
As one would expect, \AUTOGRAM is unable to parse the expressions from
\calc and \mathexpr grammars. For \cgidecode, \AUTOGRAM performed poorly,
while \Mimid achieved 100\% accuracy. For \netrc, both \AUTOGRAM and \mimid
performed poorly. However, \Mimid performed slightly better than \AUTOGRAM.
On analysis, we found that the \emph{lexing} stage introduced a large amount
of noice in the inferred parse tree, as the character access from the
\emph{lexing} stage was indistinguishable from the \emph{parsing} stage. We will
focus on this as our future work.
As we expected, \mimid performed better for \microjson too with more than 90\%
of the input samples recognized by the inferred grammar.

\begin{result}
Grammars inferred by \Mimid \emph{accept} many more correct inputs than \AUTOGRAM.
\end{result}

The outlier is \urlparse, for which \AUTOGRAM
performed achieved 100\% while \Mimid performed slightly worse (but still more
than 90\% input strings recognized by the inferred grammar). An inspection of
the source code of the subject program reveals that it violated one of the
assumptions of \mimid.  Namely, \urlparse searches for character strings in the
entirety of its input rather than restricting searches to unparsed parts of the
program. For example, it searches for URL fragments (delimited by \texttt{\#})
starting from the first location in the input. When this happens, \mimid has no
way to tell these spurious accesses apart from the true parsing.

\subsection{RQ 3. Parsers without Data Flow to variables}

To investigate whether \Mimid would also work on parsers that do not rely on data flow of input fragments to variables, we create a variant of the program in
\Cref{fig:calcprogram}. It is a simple recognizer\footnote{
A recognizer recognizes the structure but does not return a parse tree.} for calculator expressions,
and is given in Figure~\ref{fig:calcacceptor}.
\begin{figure}[H]
  \begin{lstlisting}[style=mystyle, language=Python]
    def is_digit(i):
        return i in "0123456789"

    def parse_num(s,i):
        while i != len(s) and is_digit(s[i]):
            i = i +1
        return i

    def parse_paren(s, i):
        assert s[i] == '('
        i = parse_expr(s, i+1)
        if i == len(s):
            raise (s, i)
        assert s[i] == ')'
        return i+1

    def parse_expr(s, i = 0):
        is_op = True
        while i < len(s):
            c = s[i]
            if is_digit(c):
                if not is_op: raise (s,i)
                i = parse_num(s,i)
                is_op = False
            elif c in ['+', '-', '*', '/']:
                if is_op: raise (s,i)
                is_op = True
                i = i + 1
            elif c == '(':
                if not is_op: raise (s,i)
                i = parse_paren(s, i)
                is_op = False
            elif c == ')':
                break
            else:
                raise (s,i)
        if is_op:
            raise (s,i)
        return i

    def main(arg):
        parse_expr(arg)
  \end{lstlisting}
\caption{A recognizer for mathematical expressions.}
\label{fig:calcacceptor}
\end{figure}
Now, the parser no longer stores parsed information
in any data structure.
\iflong There are no substrings being stored in any variable,
and hence no taints to look for. \fi
\Mimid can still recover
the grammar from this program, and the parse tree is exactly the same
as the one given in Figure~\ref{fig:derivationtree}.
However, \AUTOGRAM is unable to track the data flow because there is no
data flow anymore.
This means that \Mimid is able to recover grammars from
a larger class of programs than \AUTOGRAM. (While complex programs
such as recognizers can be written without data flow, it is
impossible to write complex programs such as parsers without some control flow.)
Why is this important? Not all parts of an input may be equally valid. One may
have to parse the header to identify where the body starts even if one is not
interested in the information contained in the header. One may also wish to
skip parts of the input to get to the next interesting chunk. None of these
require explicit data flow.

\begin{result}
\Mimid infers grammars from parsers \\
without data flow from input to variables.
\end{result}

\subsection{RQ 4. Advanced Parsers}

\begin{figure}[H]
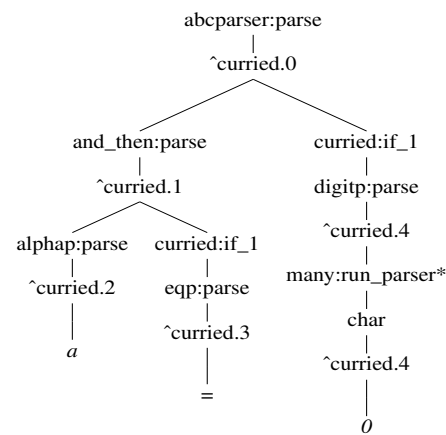

\resizebox{.70\columnwidth}{.03\totalheight}{
  \Tree[.abcparser:parse [.\^{}curried.0 [.and\_then:parse
           [.\^{}curried.1
             [.alphap:parse
                   [.\^{}curried.2 [\textit{a} ]]
              ]
            [.curried:if\_1
              [.eqp:parse
                   [.\^{}curried.3 [\textit{=} ]]
              ]
             ]
          ]
        ]
        [.curried:if\_1
         [.digitp:parse
         [.\^{}curried.4
           [.many:run\_parser*
                                 [.char [.\^{}curried.4 [\textit{0} ]]] ]
                ]
                 ]] ]]
}
  \caption{Recovered (unprocessed) parse tree from PyParsec for a simple language with top level rule:\\
  \texttt{abcparser := alphap >> eqp >> digitp}\\
  the \texttt{*} indicates where a long chain was shortened}
\label{fig:parsec}
\end{figure}

While a large number of parsers are written by hand~\cite{mathis2019parser} in the
traditional recursive descent approach, a few other parsing techniques in
the program stack based recursive descent family have
become popular recently: Parser combinators and PEG parsers.
In fact, parser combinators~\cite{chifflier2017writing} and PEG
parsers~\cite{koprowski2010trx} are recommended over parser generators
due to the various inflexibilities such as handling
ambiguities, context-sensitive features~\cite{laurent2016taming}, and bad error
messages~\cite{jeffery2003generating}\footnote{
In the words of a commenter \url{https://news.ycombinator.com/item?id=18400717}
``getting a reasonable error message out of YACC style parser generators is as
fun as poking yourself in the eye with a sharp stick''. GCC
\url{http://gcc.gnu.org/wiki/New_C_Parser}, and CLANG  \url{http://clang.llvm.org/features.              html\#unifiedparser}
uses handwritten parsers for the same reason.
Python is also shifting to a
PEG parser \url{https://medium.com/@gvanrossum_83706/peg-parsers-7ed72462f97c}.
} when using parser generators.
Our technique can recover parse trees for both kinds of parsers, which can
be used to recover grammar as we detailed previously.
For example, Figure~\ref{fig:parsec} shows the parse tree obtained from
a simple parser written using \emph{PyParsec}\footnote{
\url{https://pypi.org/project/pyparsec/}
} given below.
\begin{lstlisting}[style=python]
import pyparsec
alphap = pyparsec.char('a')
eqp = pyparsec.char('=')
digitp = pyparsec.digits
abcparser = alphap >> eqp  >> digitp
\end{lstlisting}

\iflong
(we use
the PEG parser from the Parser~\cite{fuzzingbook2019:Parser} chapter from Zeller et al.).
Using the following grammar:
\begin{grammar}
<main> ::= <assignment>

<assignment>  ::= <alpha> '='  <digit>

<alpha>  ::= `a' | `b' | `c' | `d' | `e' | `f' | `g' | `h' | `i'

<digit> ::= `0' | `1' | `2' | `3' | `4' | `5' | `6' | `7' | `8' | `9'
\end{grammar}
with PEG parser, Figure~\ref{fig:peg} details the recovered parse tree for the input: \texttt{a=0}
\fi
\iflong
\begin{figure}[H]
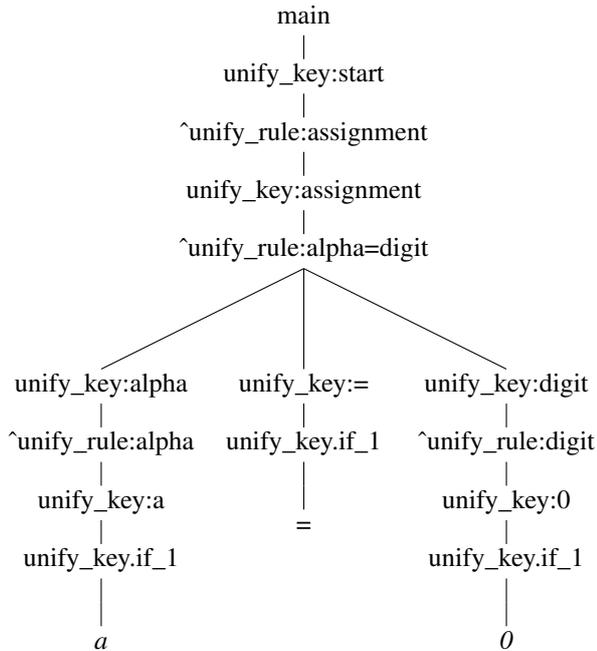

\Tree[.main [.unify\_key:start
     [.\^{}unify\_rule:assignment
       [.unify\_key:assignment
            [.\^{}unify\_rule:alpha=digit
              [.unify\_key:alpha [.\^{}unify\_rule:alpha [.unify\_key:a [.unify\_key.if\_1 [ \textit{a} ] ] ]  ] ]
              [.unify\_key:= [.unify\_key.if\_1 [ \textit{=} ]]]
              [.unify\_key:digit [.\^{}unify\_rule:digit [.unify\_key:0 [.unify\_key.if\_1  [ \textit{0} ]] ]  ] ]
            ] ] ]]]
  \caption{Recovered (unprocessed) parse trees from PEG parser for a simple assignment language with top level rule:\\
    \texttt{assignment := alpha '=' digit} on parsing \texttt{a=0}}
\label{fig:peg}
\end{figure}
\fi
Due to limitations that Python imposes, the parser
combinator library needs to be slightly modified to recover the
node names. In particular, the function names themselves do not
have any relation with the parsing behavior. Rather, chains of parser objects
are assigned to specific variables, and the names of these variables
(such as \texttt{alphap}, \texttt{eqp} and \texttt{digitp}) capture the
parse information\footnote{
Given that variables are involved, would \AUTOGRAM perform better with
parser combinators? The problem here is that the variables do not hold
parsed fragments of strings. Rather, they are functions in disguise,
and the result of parsing is not stored anywhere. Hence, \AUTOGRAM has
no usable dataflow to extract.}.
For the parser combinator library, we capture the variable names as
\nonterminal symbols in the grammar.

We can also recover the parse trees, and hence the grammar from a PEG parser,
with similar results as that of the parser combinator. For details, refer to
the Jupyter notebook submitted along with the paper.

\begin{result}
\Mimid can infer the context-free grammar \\
from PEG and combinator parsers.
\end{result}

\section{Limitations}
\label{sec:limitations}

Our work is subject to the following important limitations.

\begin{description}
\item[Approximation.]  Any approach trying to recover an input language as a context-free grammar can only produce an \emph{approximation} of the actual input language. (A fully accurate description would have to be Turing-complete just as the accepting program---i.e. an unrestricted grammar or a program accepting the input).
\iflong
When using grammars for understanding programs and input formats, this limitation is well-known, as grammars for, say, programming languages always only have covered syntactic aspects only.
\fi
    For the purpose of test generation, this inaccuracy results in unnecessary (semantically) invalid    inputs being generated; but as these would be quickly rejected by the program under test, the risk is     limited to excess resource usage.

\item[Parser Combinators.] For parser combinators, the method names
  themselves do not hold any meaning. The \nonterminal symbols have
  to be extracted from the variable names, which may require library
  specific tagging or processing.
  Same issue exists for the PEG parser. While the particular PEG parser
  we used stores the name of the \nonterminal symbol in the
  argument to the parsing procedure, this may not always be the case.
  Hence, library specific processing may be required to identify human
  readable \nonterminal names.

\item[Table-driven parsers.]  \iflong
To identify the input structure, \mimid makes use of control flow and call stack during execution.
\fi
In \emph{table-driven} parsers, control flow and stack are not explicitly encoded into the program, but an implicit part of the parser state, which \mimid could recover. However, table-driven parsers are typically generated from a given grammar, which one could simply use in the first place.

\item[Sample inputs.]  The features of grammars produced by \mimid reflect the features of the inputs it learns from: If a feature is not present in the input set, it will not be present in the resulting grammar either.
\iflong
This problem of incompleteness, which is shared with all current grammar induction techniques, can be alleviated by learning from a large and varied set of inputs.
\fi
New test generators specifically targeting input processors~\cite{mathis2019parser} could be able to create such input sets automatically.

\item[Reparsing.]  Since \mimid tracks only the \emph{last} access of a character, it can get challenged if an ad hoc parser reparses a previously parsed input.  This problem can be addressed by exploring multiple candidates for consumption and assessing the resulting grammars for their structure.
\end{description}

\section{Related Work}
\label{sec:background}

Learning the input language of a given program is not a new line of research.
However, the recent rapidly rising interest in automated test generation
and fuzzing has fueled an increase of interest in this field.

\subsection{Grammar Inference with Membership Queries}

The first class of approaches treats the program as an oracle which can be
asked whether it will accept a given string or not, and assumes little else.

We note that the seminal paper \emph{``When Won't Membership Queries Help?''}
by Angluin and Kharitonov~\cite{angluin1995when} shows that a pure black-box
approach is doomed to failure as \textbf{there cannot be a polynomial time algorithm}
in terms of the number of queries needed for recovering a context-free grammar
from membership queries alone. That is, as the grammar length increases, the
number of membership queries needed to infer the grammar grows exponentially.

In the class of query-based approaches, \GLADE~\cite{bastani2017synthesizing} is
a pure \emph{black-box} approach that produces grammatical structures.
\GLADE takes a program and a set of inputs and then constructs a series of
increasingly general languages by using generalization steps, which add
repetition, alternation, and recursive constructs to the language.
Each generalization is tested by sending synthesized inputs
to the program under test; if all inputs are accepted, the generalization is
valid.
The advantage of \GLADE over white-box approaches is that no complex analysis is
necessary or even possible; and the authors apply \GLADE to complex programs
such as PDF processors.  The disadvantage is that the \GLADE grammars cannot
make use of the program structure or its identifiers. Indeed, \GLADE makes no
claim as to recovering a readable grammar.
An inspection of the \GLADE source~\cite{gladesrc} shows that the mined
grammar is in the form of a internal data structure, which
is not easily translatable to a context-free grammar form. Indeed,
new programs have to be incorporated separately by implementing program
specific drivers.
Consequently, the grammar can't be consumed by any external tool that accepts a
context-free grammar such as parsers or grammar fuzzers, which makes it hard
to evaluate whether the ``grammar'' it produces is reasonable.

Bastani et al. shows that a Dyck language can be inferred by the \GLADE algorithm
in $O(n^4)$ time in terms of the seed length. However, we note that Dyck
languages are a small subset of context-free languages, and one needs to be
careful to generalize this result to the general class.

\begin{comment}
as the authors put it: ``we did not    design an approach for printing out the
grammars in human readable form, since we found them to be
unwieldy and incomprehensible''~\cite{bastani2019mail}.
\end{comment}

\REINAM~\cite{wu2019reinam}starts with the \GLADE approach, but improves on two
counts. First, it uses \emph{PEX}\cite~\cite{tillmann2008pex} for generating the
initial seed inputs. Next, it improves on the generalization algorithm of
\GLADE.  However, the grammar it produces is the same form as that of
\GLADE, and has the same drawbacks. Second, the symbolic execution engine
\emph{PEX} is only used for the initial seed generation. Hence as with \GLADE,
it ignores the internal structure of the program, which leaves it open to the
same limit pointed out by Angluin et al.~\cite{angluin1995when}.

\GRIMOIRE by Blazytko et al.\cite{blazytko2019usenix} is an end-to-end grey-box
fuzzer that uses the new coverage obtained by inputs to synthesize a
\emph{grammar like structure} while fuzzing.
There are two major shortcomings with the grammar like structures generated
by \GRIMOIRE. First, according to authors~\cite[Section 3, last paragraph]{blazytko2019usenix},
the grammar like structure contains a flat hierarchy, and contains a single
\nonterminal denoted by $\square$.
This \nonterminal can be expanded to any of the ``production rules'' which are
input fragments with the same \nonterminal $\square$ inserted in them, producing
gaps that can be filled in. The problem is that, real world applications often
have multiple nestings, where only a particular kind of can be inserted --- e.g
numbers, strings, etc. These kinds of structures cannot be represented by the
grammar like structure without loss of accuracy.
Second, as the grammar structure derived by \GRIMOIRE is essentially a long
list of templates, the grammar is likely to be uninterpretable by
humans.
%
%
%
%
%
%
%
%
\begin{comment}
%
Finally, the \GRIMOIRE authors
claim that they do not require ``any previous assumptions about the input
structure'', it does make use of the fact that characters such as
\texttt{`.'},
\texttt{`;'},
\texttt{`,'},
\texttt{`\textbackslash{}n'},
\texttt{`\textbackslash{}r'},
\texttt{`\textbackslash{}t'},
\texttt{`\#'},
and \texttt{` '} are common separators, and the fact that
\texttt{`()'},
\texttt{`[]'},
\texttt{`\{\}'},
\texttt{`<>'}, as well as single and double quotes act as nesting mechanisms.
\end{comment}
%

%
%
%
%

%

Learn\&Fuzz by Godefroid et al.~\cite{godefroid2017learn} automates the generation of an input grammar   using sample inputs and neural-network-based statistical machine-learning techniques.  This approach uses thousands of inputs (63,000 PDF objects) to learn from, and uses queries not only to determine            membership, but also to maximize code coverage.
Similarly to the tools above, the resulting grammars are not meant for human consumption, but show       considerable improvements for test generation of complex inputs. Another model inference tool is          \emph{PULSAR}~\cite{gascon2015pulsar} which recovers a Markov model and state machine representation of   the input. Another notable mentions include \emph{Neural byte sieve}~\cite{rajpal2017not}, and            \emph{NEUZZ}~\cite{shi2019neuzz}.

It should be noted that neither \GLADE, \REINAM, \GRIMOIRE, or Learn\&Fuzz
provide examples of inferred grammars or grammar-like structures in their
publications; all these are strictly meant as intermediate representations to be
passed to a fuzzer without assuming a human could make use of these.
Their usefulness for fuzzing without humans in the loop, however, is clearly
demonstrated.

\subsection{Using Program Analysis for Grammar Mining}

Given the limitations of a black-box approach toward learning input languages
--- any general algorithm will have exponential time behavior --- an obvious
choice is to recover information about the structure of the input language
from the parser in question.

\iflong
\else
\subsubsection{Autogram}
We start with approaches that, like ours, use \emph{program analysis} to extract
the input language.
\fi
\AUTOGRAM ~\cite{hoschele2016mining} (and later Pygmalion~\cite{gopinath2018sample})
is the approach closest to ours. \AUTOGRAM uses the program code in a \emph{dynamic},
\emph{white-box} fashion. Given a program and a set of inputs,
\AUTOGRAM uses \emph{dynamic taints} to identify the \emph{data flow} from the
input to string fragments found during execution.  These entities are associated
with corresponding method calls in the call tree, and each entity is assigned an
\emph{input interval} that specifies start and end indices of the string found
in that entity during execution.  Using a subsumption relation, these intervals
are collated into a parse tree; the grammar for that parse tree can be recovered
by recursively descending into the tree.
While \AUTOGRAM can produce very readable and usable grammars, its success depends on having data flow to track.
If parts of the input are \emph{not} stored in some variable, there is no data flow to learn from.
If the parser skips parts of the input (say, to scan over a comment), this will not result in data flow.  Conversely, data can flow into \emph{multiple variables,} causing another set of problems.
If a parser uses multiple functions, whose parameters are a buffer pointer and an index into the buffer, then each of these functions gets the entire buffer as data flow.  Such programming idioms may be less frequent in Java (the language \AUTOGRAM aims at), but in general would require expensive and difficult disambiguation.

In contrast, our approach tracks \emph{all} accesses of individual characters, no matter whether they would be stored.  Our assumption that the last function accessing a character is the \emph{consumer} of   this character (and hence parsing a \nonterminal) still produces very readable and accurate grammars.

\subsection{Recovering Parse Trees}

Lin et al.~\cite{lin2008deriving,lin2010reverse} show how to recover parse trees
from inputs using a combination of static and dynamic analysis. They observe
that the structure of input is induced by the way its parts are used during
execution, and provide two approaches for recovering bottom-up and top-down
parse trees. Similar to our approach, they construct a call tree which
contains the method calls, and crucially, the conditionals and individual
loop iterations. Next, they identify which nodes in the call tree consumes
which characters in the input string.

Their key idea is \emph{parsing point} which is the point at which they consider
a particular character to have been consumed, and they define the parsing
point of a character as the last point that the character was \emph{used}
before the parsing point of its successor. In particular, a character is used
when a value \emph{derived from it} is accessed --- that is, the input labels
are propagated through variable assignments much like taints (the labels are
propagated except during binary operations on two input related variables).

A problem with this approach is that, this approach \emph{only} considers well
written parsers in the text book
style, that consumes characters one by one before the next character is parsed.
Unfortunately, a large number of real world parsers are written in an ad hoc
style where we cannot have such a firm guarantee on the order of consumption.
For example, the Python URL parser first checks if a given URL contains any
fragment (indicated by the delimiter \texttt{\#}), and if there is, the fragment
is split from the URL. Next, in the remaining \emph{prefix}, the query string is
checked, which is indicated by the delimiter \texttt{?}, which is then separated
out from the path. Finally, the parameters that are encoded in the path using
\texttt{;} are parsed from the path left over from the above steps. This kind
of processing is by no means rare. A common pattern is to split the input
string into fields using delimiters such as commas, and then parse the
individual fields. Some programs may also parse the outer structure such as
the XML wrapper first before separately parsing the wrapped data. All this
means that the parsing points as determined by the algorithm by Lin et al.
will occur much before the actual parse. Lin et al. notes that one can't
simply use the \emph{last use} of a label as its parsing point because the
values derived from it may be accessed after the parsing phase.

\Mimid uses the same \emph{last use} strategy, but gets around this problem
by only tracking access to the original input buffer. That is, \mimid stops
tracking as soon as the input is transformed, which makes
the \mimid instrumentation lightweight, and its mined grammars accurate.

Finally, Lin et al. stop at parse trees and do not proceed further.
While they show how the function names can form the \nonterminal symbols, their
approach stops at identifying control flow nodes, and makes no attempt to either
identify compatible nodes or the iteration order, or to recover a grammar which
needs something similar to the \emph{prefix tree acceptor} algorithm to
generalize over multiple trees, each of which is needed to \emph{accurately label}
the parse tree.

If one has an \emph{accurately labeled} parse tree, grammar recovery is possible
as demonstrated by Kraft et al.~\cite{kraft2009grammar} and
Duffy et al.~\cite{duffy2007automated} who hack the \emph{GCC} to output the
parse tree after parsing, and use it to derive the grammar of C.
Zhao et al.~\cite{zhao2010program} recovers a graph grammar from the call trace,
but without any further generalization.

\subsection{Learning Finite State Models}

The idea of using dynamic traces for inferring models of the underlying software
goes back to \cite{hungar2003test}, learning a finite state model representation
of a program; Walkingshaw et al.~\cite{walkinshaw2007reverse} later refined this
approach using queries.  Such models represent legal sequences of (input)
\emph{events} and thus correspond to the input language of the program.
While our approach could also be applied to event sequences rather than
character sequences, it focuses on recovering syntactic (context-free)
input structures.

\subsection{Domain-Specific Approaches}

Some domains have seen related language recovery techniques.
Polyglot~\cite{caballero2007polyglot} by Caballero et al. and
Prospex~\cite{comparetti2009prospex} from Comparetti et al. reverse engineer
network protocols.  They track how the program under test accesses its input
data, recovering fixed-length fields, direction fields, and separators.
Tupni~\cite{cui2008tupni} from Cui et al. uses similar techniques to reverse
engineer binary file formats; for instance, element sequences are identified
from loops that process an unbounded sequence of elements.
AuthScan~\cite{bai2013authscan} from Bai et al. uses source code analysis to
extract web authentication protocols from implementations.  None of these
approaches generalizes to recursive input structures (and hence grammars),
as \mimid does; in our context, these techniques could be useful to extract
input grammars for programs processing binary inputs.

\subsection{Summary} Of all the grammar induction and mining methods
discussed, very few have actually managed to recover an actual
\emph{context-free grammar} rather than a
\emph{grammar like structure} that is opaque to practitioners. We note that
only \AUTOGRAM and \mimid are the only methods available today that
can recover a full and accurate context-free grammar.

\section{Conclusion and Future Work}
\label{sec:conclusion}

A formal input model for a program can be of use in diverse fields. However,
formal input models are often unavailable, and in cases when it is available,
the model can be incomplete, obsolete, or inaccurate with respect to the
program in the vast majority of cases.
Inferring input grammars from dynamic control flow produces readable grammars
that accurately describe input syntax.  Improving over the state of the art,
which uses data flow to identify grammars, our approach can infer grammars
even in the absence of data flow, does not require heuristics for common parsing
patterns, and is applicable to a wide range of parsers.  As the evaluation
shows, it is superior to the state of the art both in precision and recall.%

Having an algorithm that reliably produces input grammars offers several
research opportunities.  Besides addressing limitations (\Cref{sec:limitations}),
our future work will have following focus:

\begin{description}
\item[Other languages.]  Right now, our approach is implemented in Python and
works on Python files;
this has the advantage of it
being entirely self-contained, easy to assess and reproduce.
Having established its effectiveness, we are currently porting
\mimid to binary executables. This is not too hard, as both dynamic tracking of
input characters and dynamic control flow is available on binary level too.
In particular, tools such   as Ghidra\footnote{\url{https://github.com/NationalSecurityAgency/ghidra}}
can easily recover the control structure of a binary program, while tools
such as Pin\footnote{\url{https://software.intel.com/en-us/articles/pin-a-dynamic-binary-instrumentation-tool}}
can dynamically insert probes into binaries without
recompilation. Hence, while we have utilized the availability of source code
in this paper, there is nothing that prevents the same approach from being
implemented on (non-obfuscated) binaries.

\item[Tokenization.] For efficiency, input processors often consist of a
\emph{scanning} phase, composing characters into \emph{tokens}, and an actual
\emph{parsing} phase, composing tokens into syntactical structures.
We are working on automatic identification of scanners in existing programs,
applying our approach first to the scanner, and by identifying the relationship
between input elements and tokens, express grammars by means of tokens rather
than characters.

\end{description}

The complete code of our approach, including the subjects and experimental data, is available as a self- contained Jupyter notebook, which can be applied to arbitrary Python programs.  For details, go to the site
\begin{center}
\url{https://github.com/vrthra/arxiv-mimid/blob/1.0/PymimidBook.ipynb}
\end{center}

\printbibliography
\end{multicols}
\end{document}